\documentclass{llncs}

\usepackage{amssymb}
\usepackage{graphicx}
\usepackage{url}
\urldef{\mailsa}\path|{alfred.hofmann, ursula.barth, ingrid.haas, frank.holzwarth,|
\urldef{\mailsb}\path|anna.kramer, leonie.kunz, christine.reiss, nicole.sator,|
\urldef{\mailsc}\path|erika.siebert-cole, peter.strasser, lncs}@springer.com|    
\newcommand{\keywords}[1]{\par\addvspace\baselineskip
\noindent\keywordname\enspace\ignorespaces#1}

\usepackage{subfigure}
\usepackage{xcolor}

\begin{document}

\mainmatter  

\title{Diffusion of Innovations \\ over Multiplex Social Networks%
\thanks{This work has been partly funded by FIRB project Information monitoring, propagation analysis and community detection in Social Network Sites.}}

\author{Rasoul Ramezanian\inst{1}
\and Mostafa Salehi\inst{2,3}
\thanks{Corresponding author (mostafa.salehi@unibo.it)}%
\and \\Matteo Magnani\inst{4}
\and Danilo Montesi\inst{3}
\institute{ Sharif University of Technology, Iran
\and University of Tehran, Iran
\and University of Bologna, Italy
\and Uppsala University, Sweden
 }
}

\maketitle

\begin{abstract}
The ways in which an innovation (e.g., new behaviour, idea, technology, product) diffuses among people can determine its success or failure. In this paper, we address the problem of diffusion of innovations over multiplex social networks where the neighbours of a person belong to one or multiple networks (or \emph{layers}) such as friends, families, or colleagues. To this end, we generalise one of the basic game-theoretic diffusion models, called \emph{networked coordination game}, for multiplex networks. We present analytical results for this extended model and validate them through a simulation study, finding among other properties a lower bound for the success of an innovation.
While simple and leading to intuitively understandable results, to the best of our knowledge this is the first extension of a game-theoretic innovation diffusion model for multiplex networks and as such it provides a basic framework to study more sophisticated innovation dynamics.
\end{abstract}

\section{Introduction}\label{sec:intro}

How does an innovation (e.g., new behaviour, idea, technology, product) diffuse among people? The study of this problem by considering direct-benefit effects, in which decisions about adoption is based on the direct payoffs from copying the decisions of others, has received a great deal of interest in social sciences \cite{Blume1993,Morris2000,Immorlica2007,MatthewO.Jackson2005}. In this case, the success of an innovation depends on its overall compatibility with the social system that is spreading it. 

Previous studies modelled this type of diffusion processes as a coordination game played on a single social network \cite{Morris2000,Immorlica2007}. However, real diffusion phenomena are seldom constrained inside a single network. 
Suppose there are two communication systems A and B which are not interoperable (e.g, Skype and Viber). Then users must be on the same system in order to communicate. In practice, the neighbours of a user belong to different networks (also known as layers, dimensions, aspects or circles) such as friends, families, or colleagues. We refer to these as a \emph{multiplex network} \cite{Wasserman1994,Kivela2013}. Choosing one of the two aforementioned systems is based on the payoffs that arise from using compatible systems instead of incompatible ones. In addition, splitting the network into multiple layers we can model more realistic scenarios than with single networks.

For example, individuals may have a preference of being consistent inside the same layer, e.g., having the same system among all friends, but they can be fine using a different system to communicate within other layers, e.g., their colleagues. In addition, the fact of adopting a different system may be more or less problematic on different layers, e.g., being able to communicate with family or close friends may be perceived as more important than communication within other circles. In summary, we can allow different pay-offs on the different layers. While in this paper we restrict our attention to a specific well known model of innovation diffusion, our extension of this model to multiplex networks opens several possibilities to study these and other realistic diffusion models. 

In particular, here we address the problem of diffusion of innovations over multiplex social networks by focusing on the direct-benefit effects. To this end, we first generalise the game-theoretic diffusion model proposed in \cite{Morris2000} for multiplex networks. Based on the proposed model, we compute the probability of adopting an innovation by considering different parameters such as structural properties of the underlying network, number of layers, fraction of initial adopters, and payoffs in the layers.
\section{Models and Formalisms}\label{sec:anal}
In this section we introduce our generalisation of networked coordination game \cite{Morris2000} and derive the probability that a node decides to adopt an innovation in a multiplex network with $l$ layers. 

\subsection{Underlying network}
First we describe a social network as the backbone on which the multiplex coordination game is implemented. 
Let $G$ be a multiplex social network which includes a set of $l$ different layers with the same number $n$ of nodes. Indeed, the same nodes are present on all layers while a node's neighbours can be different on different  layers. Let the pair $(u,i)$ denote node $u$ in layer $L_i$ and $(u,v,i)$ indicate a link between nodes $u$ and $v$ in layer $L_i$.

Here we assume $G$ is a network with $n$ nodes and $l$ layers modelled as \emph{Erd\H{o}s-Renyi (ER)} random graphs \cite{Erdos1959} with probability $p$ of existence of randomly placed links between two nodes.

\subsection{The formulation of multiplex coordination game}\label{sec:game}
Next, we address a case in which each node will choose from two possible behaviours $A$ and $B$.
This can be modelled as a game in which users are the players and $A$ and $B$ are the possible strategies.
In particular, in an underlying multiplex social network, each node $u$ is playing a copy of the game with each of its neighbours $v$ in the layer $L_i$ at time step $t$ based on the following payoff matrix:
(i) If both $u$ and $v$ adopt $A$, they both collect payoff $a_i>0$; 
(ii) If $u$ and $v$ both adopt $B$, they both collect payoff $b_i > 0$;
(iii) If $u$ and $v$ adopt opposite behaviours, they both collect $0$.
Moreover we assume sum of payoffs is the same for all layers, i.e. $a_1+b_1=...=a_l+b_l$. This leads to a more elegant mathematical treatment and constitutes a clear entry point for subsequent extensions of our model.

After round $t$ of the game, individual $u$ collects an overall payoff $r_i^u(t)$ and $s_i^u(t)$ for adopting behaviour $A$ and $B$, respectively. The whole payoff for a player $u$ is the sum of all the payoffs collected in all layers, i.e. $r^u(t)=\sum_{i=1}^l r_i^u(t)$ for adopting behaviour $A$ (similarly, $s^u(t)=\sum_{i=1}^l s_i^u(t)$ for $B$). All nodes update their strategies simultaneously in each round. Moreover, node $u$ chooses behaviour $A$ in round $t$ if $r^u(t) > s^u(t)$. 
Let $d_i^u$ and $f_i^u$ be respectively the number and fraction of $u$'s neighbours with strategy $A$ in layer $L_i$. Then behaviour $A$ would be a better choice if 
\begin{equation}\label{thr}
\sum_{i=1}^l{f_i^u d_i^u a_i} \geq \sum_{i=1}^l{(1-f_i^u) d_i^u b_i} 
\end{equation}

The above discussion describes the full model. Time then runs forward in unit steps; in each step, each node uses
the threshold rule to decide whether to switch from $B$ to $A$. The process stops either when every node has switched to $A$, or when we reach a step where no node wants to switch. 
Suppose that everyone in the network starts with $B$ as a default behaviour. Then, a small set of \emph{initial adopters} or \emph{seeds} all decide to use $A$, and they keep using $A$.

\section{Results}\label{sec:res}

\subsection{The success of an innovation}\label{sec:prob}

Considering the game in introduced in the previous section, here we study the success of an innovation $A$, i.e. how many people adopt this new strategy. To this end, we compute the probability of adopting strategy $A$ for a given fraction of seeds with different combinations of pay-offs.  

Let $q$ be the fraction of nodes using strategy $A$, which are assumed to be distributed uniformly in the network, whereas other nodes use strategy $B$.  
Let $a_i$ be the payoff of strategy $A$ in layer $L_i$ and $b_i$ be the payoff of strategy $B$ in layer $L_i$.
For an arbitrary node $u$,  we define $P(l,q,u)$ to be the probability that node $u$ decides to switch to strategy $A$ in a multiplex network with $l$ layers when $q$ fraction of nodes are already using A. 
Let $Y_i$ be a random variable which denotes the number of nodes which use strategy $A$ and are connected to node $u$ in layer $L_i$. $Y_i$ is binomial random variable with parameters $(n-1,pq)$, where $p$ is the edge probability in the ER model. Assuming $pq$ is small and $n-1$ is large we can approximate $Y_i$ as a Poisson random variable with parameter $\lambda=(n-1)pq$.
Let $U=\sum_{i=1}^l{Y_i}$ be the sum of $l$ random variables $Y_i$. Then $U$ is a Poisson random variable with parameter $\lambda^\prime=l\lambda$. 

Moreover, since we assume all layers are generated according to an ER model with the same probability, we can say that for each node $u$, the expected number of its neighbours in all layers is the same.
Then, according to Equ.(\ref{thr}), node $u$ adopts strategy $A$ if at least
\begin{equation}\label{equ:threshold}
\beta_l={{\sum_{i=1}^l{b_i}} \over {a_1+b_1}}
\end{equation}
of its neighbours already use $A$. We call $\beta_l$ an \emph{adopting threshold}. Then $U \geq \beta_l p(n-1)$, and we have

\begin{equation}\label{equ:Pi}
P(l,q,u) \approx P(U \geq \beta_l p(n-1))=1-\sum_{i=1}^{\lfloor \beta_l p(n-1) \rfloor}e^{-l\lambda} {{{(l\lambda)}^i}\over{i!}}.
\end{equation}
where $\lambda=pq(n-1)$.
Let $q_m$ be the fraction of nodes which use $A$ at step $m$ of diffusion. We assume $q_0$  fraction of nodes are initially use $A$. Then, at step $m+1$, a fraction $q_m$ of nodes already use $A$. Among the $(1-q_m)$ fraction of nodes who do not use $A$, $P(l,q_m,u)$ decide to switch to $A$. Thus
\begin{equation}\label{equ:PiAll}
q_{m+1}=(1-q_m)P(l,q_m,u)+q_m
\end{equation}

The first $m$ in which $q_m$ is almost near 1 shows that \emph{Complete Cascade}, where every node switches from $B$ to $A$, happens at step $m$ of diffusion for threshold of $\beta_l$.
Assume that $\alpha$ is a lower bound for $P(l,x,u)$, i.e. for all $x$, $P(l,x,u)>\alpha$. Using Equ.(\ref{equ:PiAll}), we have $q_{m+1} > \alpha+(1-\alpha)q_m$. If we iterate this inequality we derive
\begin{equation}\label{equ:q_m}
q_{m+1}> \alpha(\sum_{i=1}^m (1-\alpha)^i)+ {(1-\alpha)}^{m+1} q_0
=(1-(1-\alpha)^{m+1})+(1-\alpha)^{m+1} q_0
\end{equation}
where $\alpha=P(l,q_0,u)$ which can be computed by Equ.(\ref{equ:Pi}). With the simple adoption strategy implemented in our model this equation gives us a lower bound for the success of an innovation, i.e. the fraction of nodes with strategy $A$ in a multiplex network with $l$ layers and for a given number of seeds ($q_0$) and payoffs.
This bound is indicated in Figure \ref{fig:num1} for different fractions of seeds (red line). Moreover, as shown by the blue line our simulation results for the same parameters verify our analytical findings.   

\subsection{Number of layers}\label{sec:layer}

Here we address the impact of number of layers on the diffusion of innovations (i.e., number of nodes switching to the A strategy). To this end, we first generate various coupled ER networks with different numbers of layers. 
The payoff of adopting strategy $A$ and $B$ is respectively 2 and 1 for all layers (i.e., we assume the quality of $A$ is two times better than $B$). Our results in Figure \ref{fig:layers1} show that when adopting payoff for strategy $A$ is the same in all layers (i.e., the same adopting payoff for strategy $B$), increasing the number of layers leads to lower diffusion of $A$. 

However, for large enough $a$, i.e. the value of choosing $A$ must be much higher than choosing $B$, there are some conditions in which more layers result in more diffusion. In particular, we analytically prove in the Appendix that if $({{b}\over{a+b}} p(n-1))$ is much less than 1 (i.e., for large enough $a$), $P(k,q,u)<P(j,q,u)$ 
if and only if  
\begin{equation}\label{equ:num2}
q > {{\ln({\sqrt[j-k]{({{k} \over {j}})}^i})} \over {(n-1)p}}  \hspace{3mm}\forall  i<{\lfloor {{kb}\over{a+b}} p(n-1) \rfloor}
\end{equation}
where $q$ is the fraction of seeds. While this result is valid only under the assumptions introduced by our model, which do not necessarily match real social systems, it provides an example of how this model can be practically useful to understand the effects of specific diffusion patterns.

\subsection{Number of links}\label{sec:edgeProb}

\begin{figure*}[h!]
\centering
{
\subfigure[
]{\includegraphics[trim = 20mm 50mm 1mm 40mm, scale=0.3]{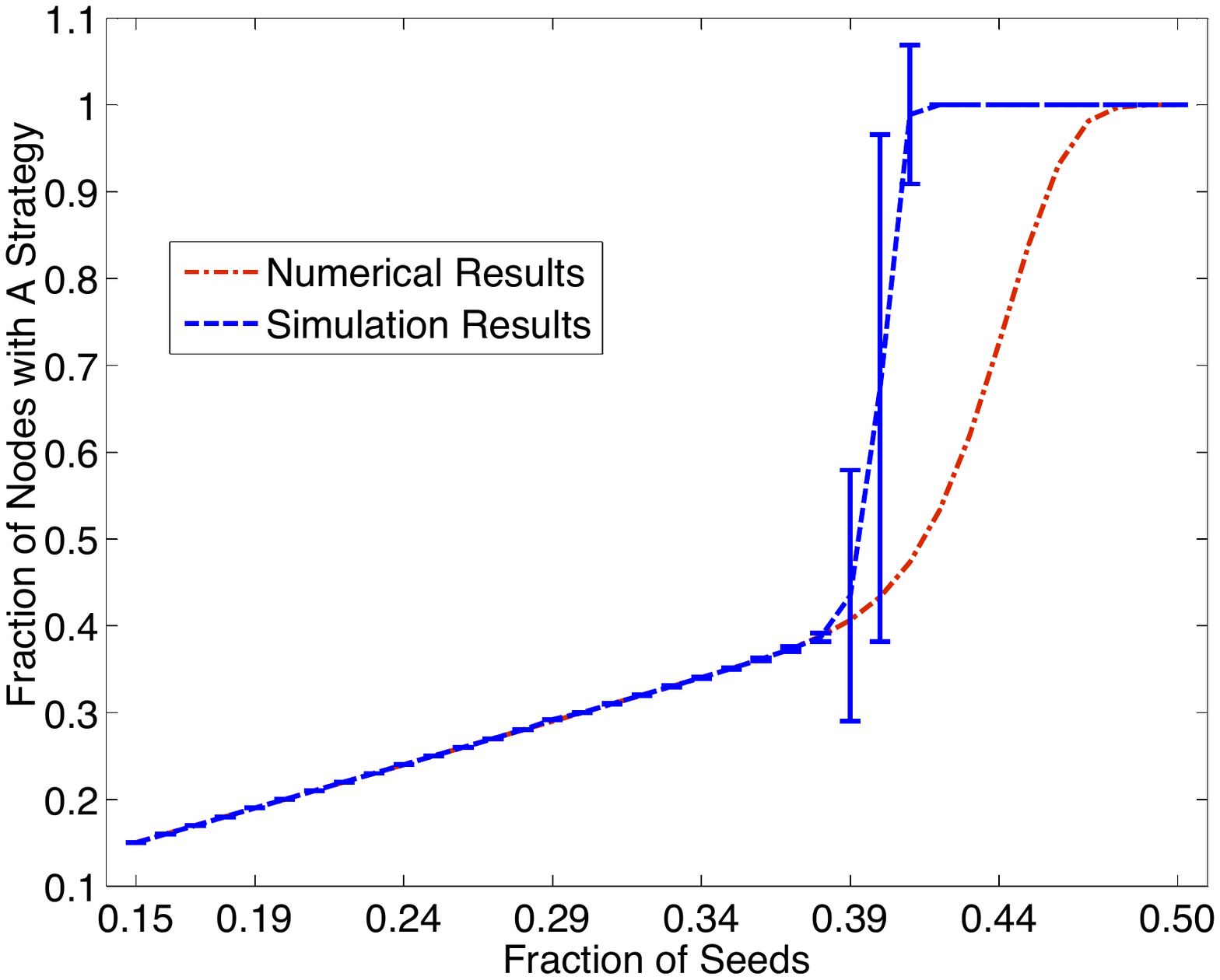}
\label{fig:num1}}
\subfigure[
]{\includegraphics[trim = 20mm 50mm 1mm 40mm, scale=0.3]{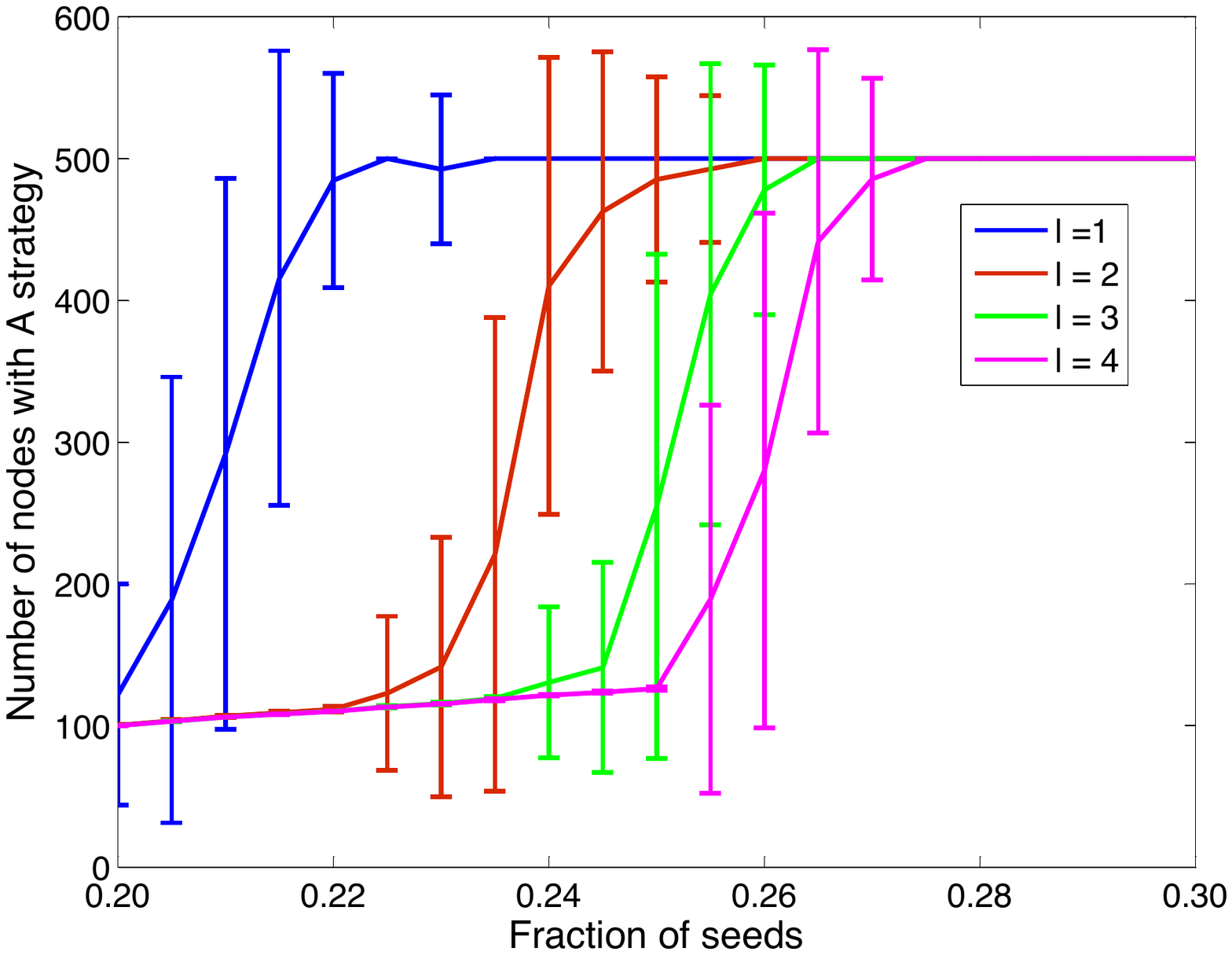}
\label{fig:layers1}}
\subfigure[
]{\includegraphics[trim = 20mm 50mm 1mm 40mm, scale=0.3]{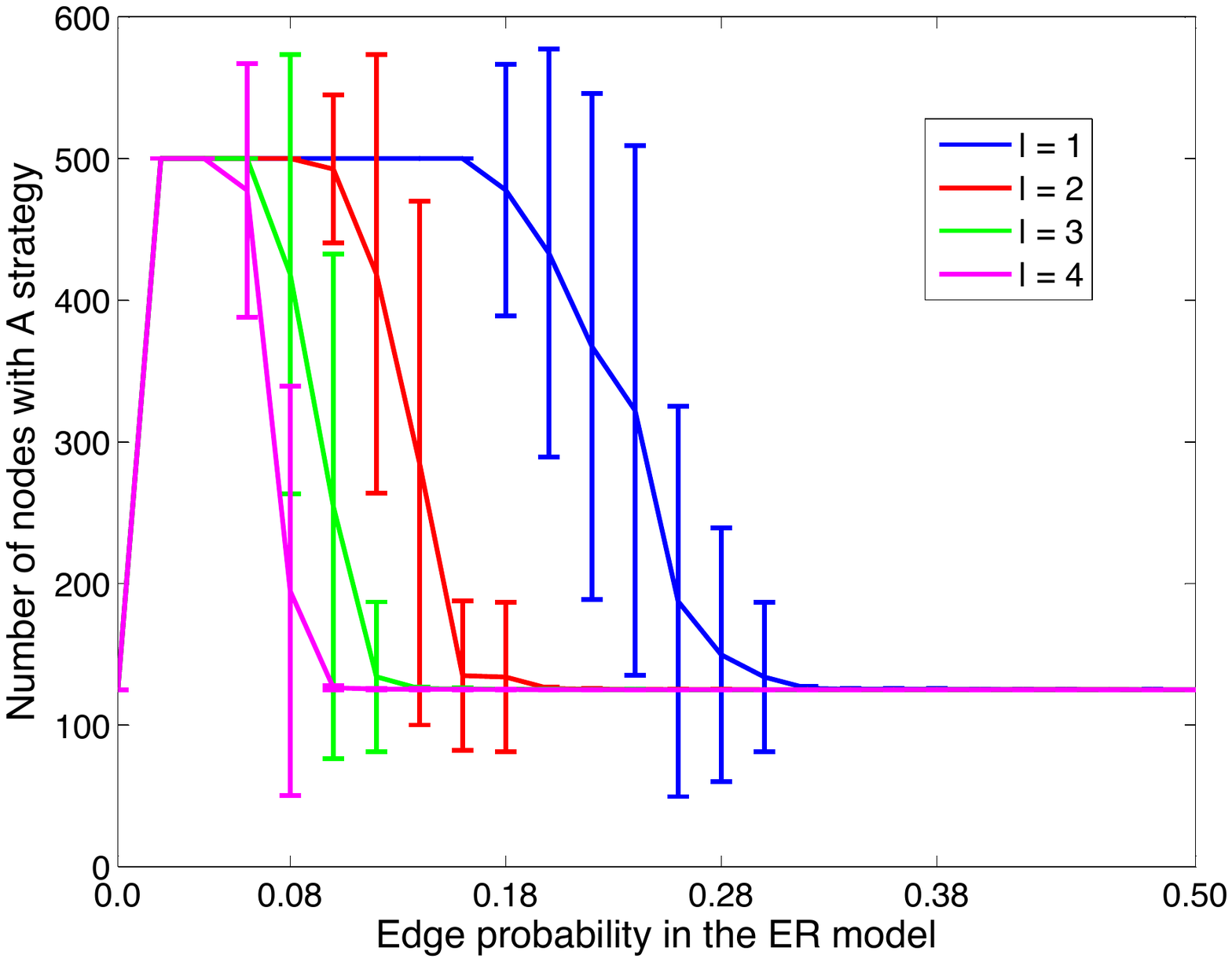}
\label{fig:edgeProb2}}
\subfigure[
]{\includegraphics[trim = 20mm 50mm 1mm 40mm, scale=0.3]{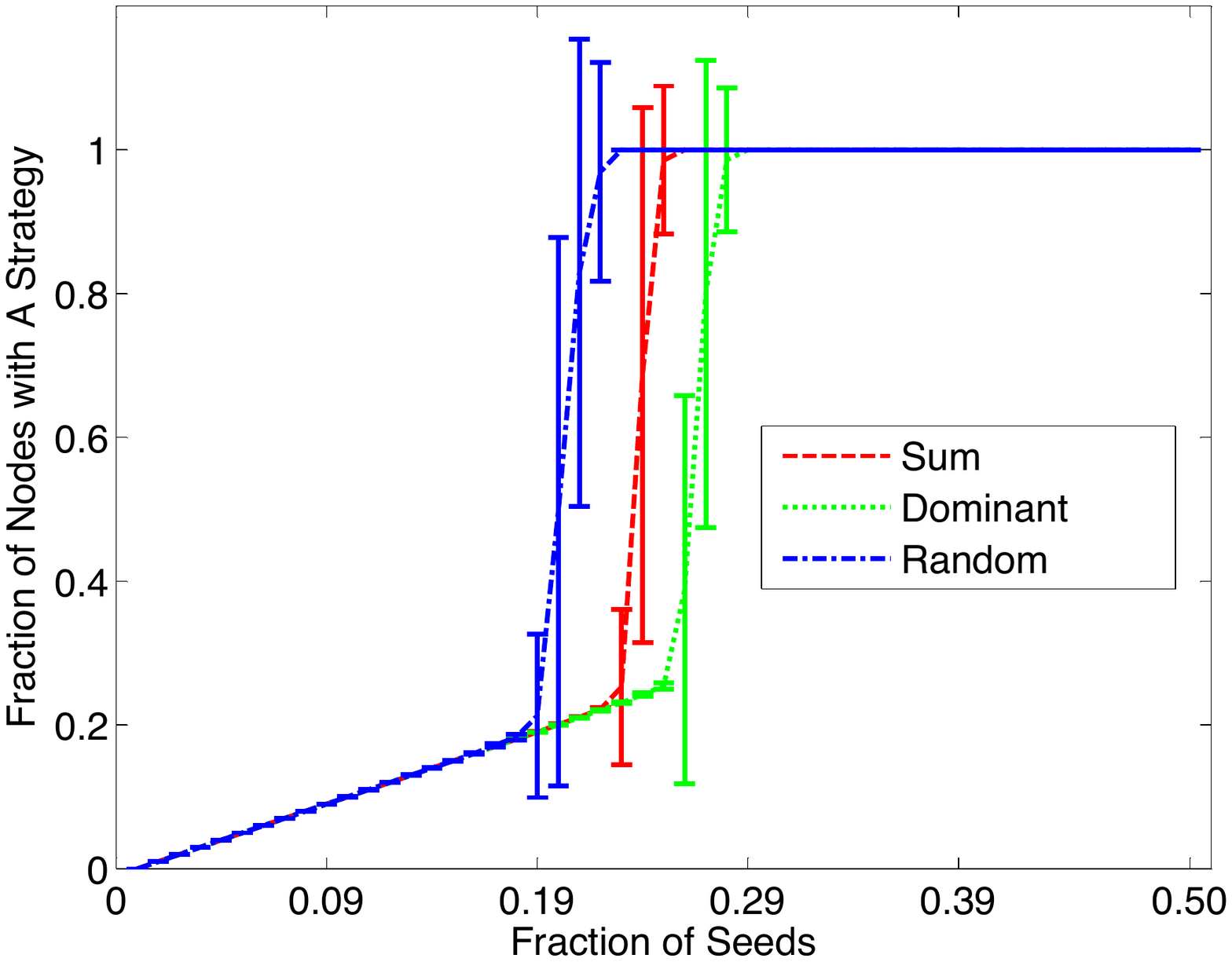}
\label{fig:approach}}
}
\caption{Simulation parameters: number of iterations= 50, Network includes ER layers, for each of them we have p (the probability of each edge), n (number of nodes). The payoffs of game: a (adopting strategy A), b (adopting strategy B).
(a) The fraction of nodes with strategy A in a two-layer multiplex network computed by Equ.(\ref{equ:q_m}) (red line) compared with simulation results with same parameters (p=0.1, n=500, a=2, b=1), (b) The impact of number of layers (p=0.1, n=500, a=2, b=1), (c) The impact of edge probabilities (p=[0,0.5], n=500, a=2, b=1, fraction of seeds=0.25). (d) The impact of different ideas for making a decision based on the payoffs (two-layer networks, p=0.1, n=500, a=2, b=1). 
}
\label{fig:fig2}
\end{figure*}

The results of our study on the impact of edge probabilities are shown in Figure \ref{fig:edgeProb2}. Increasing edge probability leads to adding new edges between nodes in the underlying multiplex network. Then, the results show the role of these new edges in shaping the diffusion of innovations over networks, i.e. how we can control diffusion by adding or removing edges from the underlying network. 
   
In general, when the adopting payoff is the same in all layers (i.e., $a=2$ and $b=1$), we observe four different phases by increasing the edge probability of the underlying network: 
(i) \emph{Adopting the new behaviour:} increasing number of nodes with strategy $A$ from number of seeds (i.e., 125 in the figure) to number of network' nodes (i.e., 500 in the figure). In this phase, both strategies $A$ and $B$ are used by the nodes;  
(ii) \emph{Epidemics of new behaviour $A$:} whole population adopts the new behaviour $A$;
(iii) \emph{Backing to old behaviour $B$:} decreasing number of nodes with strategy $A$ from number of network' nodes to number of seeds. In this phase, both strategies $A$ and B are used by the nodes;  
(iv) \emph{Epidemics of old behaviour $B$:} the whole population (except the seeds) adopts the old behaviour $B$.

Moreover, increasing the number of layers leads to a shorter phase (ii), i.e. in a shorter period all the population chooses the new behaviour $A$.

\subsection{Other decision-making strategies}\label{sec:approach}

So far we considered the sum of all the payoffs collected in all layers for making a decision to adopt a new behaviour (refer to Section \ref{sec:game} for details). We call this \emph{Sum approach}.
To highlight the value of the basic model introduced in this paper, in particular the fact that it can be easily extended to study the effects of different diffusion strategies, we compare previous results with the ones obtained using two different decision strategies:
(i) \emph{Dominant approach:} for each layer $L_i$, after round $t$ of the game, individual $u$ collects an overall payoff $r_i^u(t)$ and $s_i^u(t)$ for adopting behaviour $A$ and $B$, respectively. In this approach, node $u$ chooses behaviour $A$ in round $t$ if $r_i^u(t) > s_i^u(t)$ for all $i=1..l$. 
(ii) \emph{Random approach:}
in this approach, node $u$ randomly chooses one layer, e.g. layer $L_i$ is selected, then it adopts behaviour $A$ if $r_i^u(t) > s_i^u(t)$. 

The comparison is shown in Figure \ref{fig:approach}. As we can see the original approach lies in-between the two other approaches.

\section{Related Work}\label{sec:related}

The study of multiplex networks is a rapidly evolving research area with many challenging research issues. For a survey on different aspects of such networks we refer to \cite{Kivela2013} and references therein. A comprehensive review of diffusion processes over such networks is available in \cite{Salehi2014}. 

Although various types of diffusion processes share some common aspects, there are some differences between them. 
In summary, 
we can categorize existing studies on diffusion processes in two groups \cite{Salehi2014}: (1) epidemic-like processes and (2) decision-based processes. 
In the studies of epidemic-like processes, which are focused on disease and rumour spreading, the probability of a node getting infected by a diffusion process is determined by its neighbours \cite{Bailey1975,Hethcote2000}. 

On the other hand, studies of decision-based processes, which are also called threshold models, are based on the idea that the decision to adopt a particular idea or product by a user depends on a proportion of his/her neighbors who have already adopted it \cite{Granovetter1978,Watts2002,Centola2007a,Morris2000}. 
Moreover, existing decision-based studies follow two different approaches \cite{Easley2010}:
(i) informational effects, in which making decision is based on the indirect information about the decisions of others \cite{Watts2002}; and (ii) direct-benefit effects, in which there are direct payoffs from copying the decisions of others \cite{Kossinets2008}.

For the first approach there are some well-know models such as the Linear Threshold Model (LTM) \cite{Granovetter1978} which has also been generalised for multiplex networks in \cite{PhysRevE.85.045102,yagan-ComCotRanMulNw,MulLayNetInfProOvrMic}. Previous studies following the second approach and addressing diffusion processes modelled as a networked coordination game have been limited to a single social network so far \cite{Morris2000,Immorlica2007}. To the best of our knowledge our work is the first generalising this game for multiplex networks. 

At the same time, game theory has been used in other recent works in the field of multiplex networks. In \cite{Jiang2013,gomezPRE2012,gomez2012evolution}, the authors address how cooperative behaviour can emerge and evolve in a network of selfish nodes. Moreover, \cite{Kostka2008,Goyal2012} study the problem of interacting diffusion processes in which the diffusion dynamics of multiple concurrent and interacting processes is addressed. \cite{Kostka2008} has addressed the diffusion of competing rumours in social networks. By modelling the selection of starting nodes for the rumours as a strategic game, it has been shown that being the first player to decide is not always an advantage. \cite{Goyal2012} has studied the competition between firms who use their resources to maximize the adoption of their products.


\section{Conclusion}\label{sec:conclusion}

We studied the problem of diffusion of innovations over multiplex social networks by focusing on the direct-benefit effects. To this end, we proposed a multiplex networked coordination game by generalizing a well-known diffusion model for such type of diffusion processes. To exemplify the utility of the model to study diffusion processes we obtained a lower bound for the ratio of nodes who adopt an innovation in time (i.e., during subsequent steps of a discrete diffusion process). In particular, given some parameters such as structural properties of the underlying network, number of layers, fraction of initial adopters, and the value of innovation compared to old behaviour, one can estimate the chance of success for spreading in a population modelled as a multiplex social networks.
 

\section{Appendix}\label{app}

Assume the payoff of adopting strategy A is the same in all layers (i.e., $a_i=a \hspace{2mm} \forall i=1..l$). Since we assume $a_1+b_1=...=a_l+b_l$, then the payoff of adopting strategy B is the same in all layers (i.e., $b_i=b \hspace{2mm} \forall i=1..l$). Thus, the adopting threshold for $l$ layers will be $\beta_l={{lb}\over{a+b}}$. Considering Equ.(\ref{equ:Pi}), we have $P(k,q,u)<P(j,q,u)$ iff and only if  
\begin{equation}\label{equ:num1}
\sum_{i=0}^{\lfloor {{kb}\over{a+b}} p(n-1) \rfloor} e^{-k\lambda} {{{(k\lambda)}^i}\over{i!}} > \sum_{i=0}^{\lfloor {{jb}\over{a+b}} p(n-1) \rfloor} e^{-j\lambda} {{{(j\lambda)}^i}\over{i!}}
\end{equation}
where $k$ and $j$ denotes the number of layers.
Let  $k$ and $j$ be such that  ${\lfloor {{kb}\over{a+b}} p(n-1) \rfloor}={\lfloor {{jb}\over{a+b}} p(n-1) \rfloor}$ (if ${{b}\over{a+b}} p(n-1)$ is much less than 1, then there exist  $k$ and $j$ such that  ${\lfloor {{kb}\over{a+b}} p(n-1) \rfloor}={\lfloor {{jb}\over{a+b}} p(n-1) \rfloor}$) then inequality (\ref{equ:num1}) holds if and only if $k^i {(e^{-\lambda})}^k > j^i {(e^{-\lambda})}^j$ for all $i<{\lfloor {{kb}\over{a+b}} p(n-1) \rfloor}$.  

Consider two functions $f(x)=k^i x^k$ and $g(x)=j^i x^j$. We have for all 
$x < \sqrt[j-k]{({{k} \over {j}})}^i$, $f(x)>g(x)$. If $e^{-\lambda} < \sqrt[j-k]{({{k} \over {j}})}^i$ for all $i<{\lfloor {{kb}\over{a+b}} p(n-1) \rfloor}$ then 
$k^i {(e^{-\lambda})}^k > j^i {(e^{-\lambda})}^j$. By substituting $\lambda=(n-1)pq$,  
the inequality (\ref{equ:num1}) holds true if and only if  
\begin{equation}\label{equ:num2}
q > {{\ln({\sqrt[j-k]{({{k} \over {j}})}^i})} \over {(n-1)p}}  \hspace{3mm}\forall  i<{\lfloor {{kb}\over{a+b}} p(n-1) \rfloor}
\end{equation}
where $q$ is the fraction of seeds. For example let $n=300$, $p=0.1$, $b=1$,$a=100$. Then for $k=10$,$j=13$, we have ${\lfloor {{kb}\over{a+b}} p(n-1) \rfloor}={\lfloor {{jb}\over{a+b}} p(n-1) \rfloor}=3$.  Then if  
$e^{-\lambda} < \sqrt[3] {({{10} \over {13}})}^3$ we have $P(10, q, u) < P(13, q, u)$.

\bibliographystyle{plain}
\bibliography{mendeley}

\end{document}